\documentclass{article}
\usepackage{graphicx}
\usepackage{epsfig}

\begin{document}

\title{QED effective action for an O(2)$\times$O(3) symmetric field in the full mass range}

\author{
N. Ahmadiniaz, A. Huet, A. Raya and C. Schubert\\
\\
\it {Instituto de F\'{\i}sica y Matem\'aticas}
\\
\it {Universidad Michoacana de San Nicol\'as de Hidalgo}\\
\it {Apdo. Postal 2-82}\\
\it {C.P. 58040, Morelia, Michoac\'an, M\'exico}\\
\\
Poster presented by N. Ahmadiniaz at the \\ 
\it{XV Mexican School on Particles and Fields}\\
\it{September 6-15, 2012}\\
\it{Puebla, Mexico }\\
}
\maketitle
\baselineskip=11.6pt

\begin{abstract}
An interesting class of background field configurations in QED are the $O(2)\times O(3)$ symmetric fields.
Those backgrounds have some instanton-like properties and yield a one-loop effective action that is highly nontrivial but amenable to numerical
calculation, for both scalar and spinor QED. 
Here we report on an application of the recently developed 
``partial-wave-cutoff method'' to the numerical analysis of both effective 
actions in the full mass range. In particular, at large mass we are able to match the asymptotic behavior of the physically renormalized effective action against the leading
two mass levels of the inverse mass (or heat kernel) expansion. At small mass  we obtain good numerical results even in the massless case 
for the appropriately (unphysically) renormalized effective action after the removal of the chiral anomaly term through a small radial cutoff factor. 
In particular, we show that the effective action after this removal remains finite in the massless limit, which also 
provides indirect support for M. Fry's hypothesis that the QED effective action in this limit is dominated by the chiral anomaly term.
\end{abstract}

\section{The $O(2)\times O(3)$ symmetric background}
G. Dunne {\em et al} \cite{Dunne:2004sx, Dunne:2005te, Dunne:2006ac,Dunne:2011fp} initiated the application of 
the ``partial wave-cutoff method'', to be explained below, to the important class of $O(2)\times O(3)$ symmetric fields first introduced by 
S. L. Adler \cite{Adler:1972qq,Adler:1974nd}. We will work in Euclidean metric with 
\begin{equation}
A_{\mu} (x) = \eta^3_{\mu \nu} x_\nu g(r) \; ,\quad
g (r) \equiv \nu\,  \frac{e^{-\alpha r^2}}{\rho^2 + r^2} \;,
\label{defbackground}
\end{equation}
where $\eta_{\mu\nu}^3$ is a 't Hooft symbol, $r^2=x_{\mu}x^{\mu}$ and $\alpha\geq 0$.
In general, $g(r)$ may be any arbitrary  spherically symmetric function. 
However, the profile we have chosen for $g(r)$ has the following important properties :
\begin{itemize}
\item[(i)]  {$\alpha>0\rightarrow\int d^4x\,F^2$ is finite}.
\item[(ii)] {$\alpha=0\rightarrow g(r)\propto\frac{1}{r^2}$, which is what we need to see the chiral anomaly term $\int d^4x F_{\mu\nu}\tilde{F}_{\mu\nu}$}.
\end{itemize}
According to M. Fry \cite{fry2006,fry2010}, the following general remarks hold for the spinor QED effective action in the
background (\ref{defbackground}) with $\alpha =0$:
let $\cal R$ denote the (scheme independent) effective action obtained after subtraction of the two-point contribution.
It behaves for small $m$ as
\begin{equation}
{\cal R} \, {\sim} \, \frac{\nu^2}{4} \ln m^2 + {\rm less\,}{\rm singular}\,{\rm in}\, m^2\;.
\label{limfry}
\end{equation}
The logarithmic term is determined entirely by the chiral anomaly, 
\begin{equation}
-\frac{1}{4\pi^2}\int d^4x F_{\mu\nu}\tilde{F}_{\mu\nu}=\frac{\nu^2}{2}.
\end{equation}

\section{The partial wave-cutoff method}
After decomposing the negative chirality part of the Dirac operator into partial-wave radial operators with 
quantum numbers $l$ and $l_3$, 
the corresponding effective action is :
\begin{equation} \label{lowsum}
\Gamma^{(-)}_{{\rm{L}}} = -\sum_{s = \pm \frac{1}{2}} \; \sum_{l =
0, \frac{1}{2} , 1, \ldots}^L \; \Omega(l) \sum_{l_3 = -l}^{l}
\ln \left( \frac{ {\rm{det}}(m^2+\mathcal{H}_{(l, l_3,
s)})}{{\rm{det}}(m^2+\mathcal{H}_{(l, l_3, s)}^{{\rm{free}}})} \right) \, .
\end{equation} 
We concentrate on the negative chirality sector
of the spinor effective action 
where $
\Omega(l) =  (2 l +1) $
is  the degeneracy factor, and the
$s$ sum comes from adding the contributions of each spinor
component.
The partial-wave cutoff method separates
 the sum over the quantum number $l$ into a low partial-wave contribution, each term of which is computed using the (numerical) Gel'fand-Yaglom method, and a high partial-wave contribution, whose sum is computed analytically using WKB. 
Then we apply a regularization and renormalization procedure and combine these two contributions to yield the finite and renormalized effective 
action. 
The Gel'fand-Yaglom method \cite{Dunne:2004sx,Dunne:2005te,Dunne:2006ac}, can be summarized as follows: 
  Let $\mathcal{M}_1$ and $\mathcal{M}_2$ denote two second-order radial
differential operators on the interval $r \: \in \, [\, 0,\infty)$
and let $\Phi_1(r)$ and $\Phi_2(r)$ be solutions to the 
initial value problem 
\begin{equation}
\mathcal{M}_i \Phi_i(r) = 0; \quad \Phi_i(r) \sim r^{2 l} \quad {\rm{as}} \quad r \to 0 \, .
\label{initial}
\end{equation}
Then the ratio of the determinants is given by
\begin{eqnarray*}
\frac{{\rm{det}} \mathcal{M}_1}{{\rm{det}} \mathcal{M}_2} &=& \lim_{ R \to \infty} \left(  \frac{\Phi_1(R)}{\Phi_2(R)}      \right) \, .
\end{eqnarray*}
In our case
\begin{eqnarray}
\Phi_{-}''(r) + \frac{4 l + 3 }{r} \Phi_{-}'- \left( m^2 +
4 l_3 g(r) + r^2 g(r)^2 + [4 g(r) + r g'(r)]   \right)
\Phi_{-}(r) &=& 0 \, . \nonumber
\end{eqnarray}
The high-mode contribution, which remains to be calculated calculated using WKB, is
\begin{eqnarray} \label{gammalow}
\Gamma^{(-)}_{{\rm{H}}} &=& -\sum_{s = \pm \frac{1}{2}} \; \sum_{l = L
+ \frac{1}{2}}^\infty \; \Omega(l) \sum_{l_3 = -l}^{l} \ln
\left( \frac{ {\rm{det}}(m^2+\mathcal{H}_{(l, l_3, s)})}{{\rm{det}}(m^2+\mathcal{H}_{(l, l_3, s)}^{{\rm{free}}} )}
\right) \, .
\end{eqnarray}
\section{Two versions of the effective action}
\begin{figure}
\begin{center}
\epsfig{file=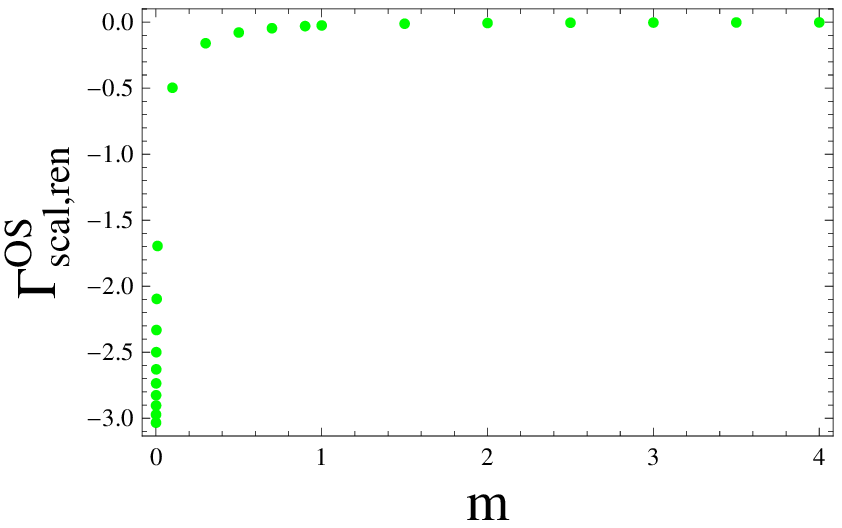,width=0.4\linewidth}
\epsfig{file=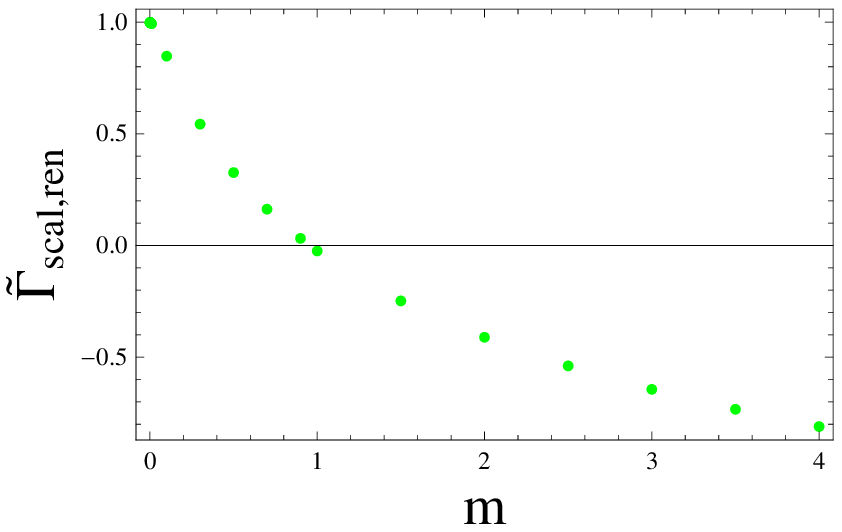,width=0.4\linewidth}
\end{center}
\caption{Effective action. {\em Left panel}, On shell, eq.~(\ref{OS}). {\em Right panel}, modified, eq.~(\ref{defGammatilde}).}
\label{fig1}
\end{figure}

For the class of backgrounds considered here, the partial-wave-cutoff  method works well for any value of the mass up to numerical accuracy.
The effective action calculated as above is finite for any non-zero value of the mass. When we use on-shell (`OS') renormalization ($\mu = m $),
its leading small-mass behavior contains the logarithmically divergent term \cite{Dunne:2011fp} 
\begin{equation}
\Gamma_{\rm{ren}}^{\rm OS} (m) \sim \Bigg( - \int_0^\infty dr \, Q_{{\rm {log}}}(r) \Bigg) \ln m \, , \quad \quad   m \to 0 \, .
\label{OS}
\end{equation}
Thus for the study of this small $m$ regime we introduce a modified effective action,
\begin{equation}
\tilde{\Gamma}_{\rm{ren}} (m) \equiv \Gamma_{\rm{ren}} (m,\mu) + \Bigg(  \int_0^\infty dr \, Q_{{\rm {log}}}(r) \Bigg) \ln \mu \quad
\Bigl( =  \Gamma_{\rm ren}(m,\mu =1) \Bigr).
\label{defGammatilde}
\end{equation}
It turns out that $\tilde{\Gamma}$ is finite for $m = 0$, 
which supports Fry's conjecture, mentioned above, for the case of the backgrounds with $\alpha > 0$ (where the chiral anomaly term is absent).
In Fig.~\ref{fig1} we contrast both variants of the effective action for the 
Scalar QED case (see \cite{NAAC} for the fermionic case which is very similar).

\section{Large mass asymptotic behavior}
\begin{figure}
\begin{center}
\epsfig{file=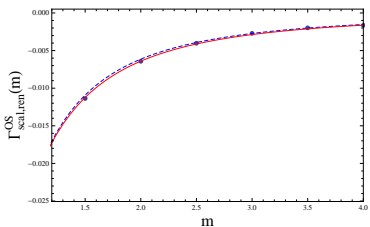,width=0.4\linewidth}
\end{center}
\caption{Large mass limit of the scalar effective action for $\alpha=1/120$. Dots
represent the exact effective action, the dashed curve considers only the leading term
$c_2/m^2$, and the solid curve, both the leading and subleading terms  $c_2/m^2 + c_4/m^4$.}
\label{fig2}
\end{figure}

 In this section we exhibit the leading and subleading terms in the inverse mass (= heat kernel) expansion of the
one-loop scalar QED effective action.
The first two terms 
are (we calculated them using the worldline formalism along the lines of \cite{5,41})
\begin{equation} \label{gammafitsp}
\Gamma_{{\rm scal}}^{\rm OS}(m) = \frac{c_{{\rm scal},2}(\alpha)}{m^2}+ \frac{c_{{\rm scal},4}(\alpha)}{m^4} + O\left(\frac{1}{m^6}\right),
\end{equation}
where the coefficients in the limit $\alpha\rightarrow 0$ are, up to cubic order in $\alpha$,
\begin{eqnarray}
c_{{\rm scal},2}(\alpha)&=& -\frac{1}{460} \alpha ^3 \left(256 \log (2 \alpha )+256 \gamma_E +\frac{304}{5}\right)-\frac{31 \alpha ^2}{300}+\frac{23 \alpha }{600}-\frac{2}{75}\;, \nonumber\\
c_{{\rm scal},4}(\alpha)&=&
-\frac{37 \gamma_E  \alpha ^3}{135}+\frac{187349 \alpha ^3}{396900}-\frac{2\gamma_E  \alpha ^2}{15}-\frac{36853 \alpha ^2}{529200}-\frac{1}{135} (37 \alpha +18)
   \alpha ^2 \log (2 \alpha )\nonumber\\
   &&-\frac{571 \alpha }{22050}+\frac{107}{52920}\;.
   \end{eqnarray}
The large-mass behavior of the effective action is shown in Fig.~\ref{fig2} for the scalar QED case (see \cite{NAAC} for the fermionic case).

\section{Finiteness of the massless four-point contribution}
 In this section we show that the four-point contribution to 
the effective action in the ``standard''  $O(2)\times O(3)$ symmetric background,
(\ref{defbackground}) with $\alpha =0$ and $\nu = \rho =1$,
 is finite in the massless limit. This is a detail of some importance for Fry's investigation that had been missing in the 
analysis of \cite{fry2006}, although it has been anticipated in \cite{fry2010}. 
In the worldline formalism, we can write this quartic contribution to the effective action  as (in either scalar or spinor QED)
\begin{eqnarray}
\Gamma^{(4)}[A] = -\prod_{i=1}^4
\int \frac{d^4k_i}{(2\pi)^4}\bar a(k_i^2)
(2\pi)^4\delta^4(\sum k_i)
\Gamma[k_1,\varepsilon_1;\cdots;k_4,\varepsilon_4]\;,
\label{Gamma4fin}
\end{eqnarray}
where $\Gamma$ is the worldline path integral representation of the 
off-shell Euclidean four-photon amplitude and $
\bar a(k^2) = 4c\pi^2 K_2(\rho\sqrt{k^2})/k^2,$
where $K_2(x)$ is the modified Bessel function of the second kind.
After performing the path integral, suitable integrations by parts, a rescaling $\tau_i = Tu_i,i=1,\ldots,4$
and the elimination of the global $T$ integral, we obtain (see \cite{41} for details) 
\begin{eqnarray}
\Gamma[k_1,\varepsilon_1;\cdots;k_4,\varepsilon_4]
=
-\frac{e^4}{(4\pi)^2} 
\int_0^1 du_1du_2 du_3 du_4\
\frac{Q_4(\dot G_{B12},\ldots,\dot G_{B34})}
{\Bigl(m^2 -\frac{1}{2} \sum_{i,j=1}^4 G_{Bij}k_i\cdot k_j\Bigr)^2}\;.
\label{4photfin}
\end{eqnarray}
Here 
$G_{Bij} \equiv G_{B}(u_{i},u_{j})= |u_i-u_j|- (u_i-u_j)^{2}$
is the worldline Green's function and $\dot G_{Bij} $
its derivative. $Q_4$ is a polynomial in the various $\dot G_{Bij}$'s, as well as in the 
momenta and polarizations. 
Now, the QED Ward identity implies that the rhs of (\ref{4photfin}) is $O(k_i)$ in each of the four momenta, which can
also be easily verified using properties of the numerator polynomial $Q_4$.
Using this fact and  (\ref{Gamma4fin}) we see that there is no singularity at  $k_i=0$, and convergence at large $k_i$.

\section{Conclusions}
We have continued and extended here the full mass range analysis of the scalar and spinor QED effective actions
for the $O(2)\times O(3)$ symmetric backgrounds, started in \cite{Dunne:2011fp}, by a more detailed
numerical study of both the small and large mass behaviors.
In \cite{Dunne:2011fp} only the unphysically renormalized versions $\tilde\Gamma_{\rm ren}(m)$ of these effective actions
were considered (corresponding to $\mu=1$), which are appropriate for the small mass limit, but have a logarithmic 
divergence in $m$ in the large $m$ limit.
Here we have instead used the physically renormalized effective actions $\Gamma_{\rm ren}^{\rm OS}(m)$ 
for the study of the large mass expansions, which made it possible to achieve a numerical matching of both this
leading and even the  subleading term in the inverse mass expansions of the effective actions.
In our study of the small mass limit, we have improved on  \cite{Dunne:2011fp} by obtaining good numerical results
for $\tilde\Gamma_{\rm ren}(m)$ even at $m=0$, and showing continuity for $m\to 0$  for various values of $\alpha$.
Moreover, we have presented numerical evidence that $\tilde\Gamma_{\rm ren}(m=0)$ stays finite even in the limit
$\alpha\to 0$. This fact is important in the spinor case, where it supports indirectly Fry's conjecture \cite{fry2006} that, for the case at
hand,  the only source of a divergence of $\tilde\Gamma_{\rm ren}(m)$ for $\alpha=0$ at $m\to 0$ should be 
the chiral anomaly term. 
As a side result, we have proved the finiteness of the massless limit four-point contribution to the effective action in scalar and spinor QED for the 
standard $O(2)\times O(3)$ symmetric background  ($\alpha=0$, $\nu =\rho =1$).\\
More details and results for the spinor QED case will be given in a forthcoming publication \cite{NAAC}.

\end{document}